\documentclass[aps,twocolumn,showpacs,showkeys,amsmath,amssymb,superscriptaddress,nofootinbib,floatfix,longbibliography,noeprint]{revtex4-2}
\usepackage{amsmath,graphicx,color,ulem}
\usepackage{epsfig,bm,slashed,hyperref}

\begin{document}

\title{Magnetic properties of the hadron resonance gas with physical magnetic moments}

\author{Rupam Samanta}
\email{rupam.samanta@ifj.edu.pl}
\affiliation{Institute of Nuclear Physics, Polish Academy of Sciences,  31-342 Cracow, Poland}

\author{Wojciech Broniowski}
\email{wojciech.broniowski@ifj.edu.pl}
\affiliation{Institute of Nuclear Physics, Polish Academy of Sciences,  31-342 Cracow, Poland}

\begin{abstract} 
We study magnetic properties of the Hadron Resonance Gas in the presence of a strong ($0 \le B \le 0.15~{\rm GeV}^2$) magnetic field, 
using physical values of the magnetic moments of hadrons, i.e., including their anomalous parts. The values of these moments are taken from 
experiment, or when unavailable, from theoretical estimates. We evaluate the conserved charge susceptibilities, 
finding the expected sizable effects of the anomalous magnetic moments, in particular for the octet baryons, such as the proton and neutron, where they are exceptionally large. We also study in detail the large effects of the magnetic moments of the $\Delta(1232)$ states, for which various theoretical 
estimates and experimental values differ significantly. 
We compare our model results with the lattice QCD data and find reasonable agreement within the model uncertainty. 
\end{abstract}

\maketitle

\section{Introduction}

The behavior of the hadronic medium under strong magnetic fields ($B$) has been a topic of active research, due to its relevance in 
hot systems formed in the early stages of ultra-relativistic nuclear collisions as well as in the early universe 
(for reviews, see, e.g.,~\cite{Kharzeev:2012ph,Andersen:2014xxa,Huang:2015oca,Adhikari:2024bfa}),
or in the cold neutron star matter (for a review, see, e.g.,~\cite{Harding:2006qn}). 
On the lattice~\cite{Bali:2011qj,HotQCD:2012fhj,Endrodi:2013cs,Bali:2014kia,Bellwied:2019pxh,Bali:2020bcn,Bollweg:2021vqf,Ding:2023bft,Ding:2025jfz}, 
one can study the effects of $B$ in a thermally equilibrated hot QCD medium in a static setup, where $B$ is spatially uniform and time-independent (unlike in the collisions, where the system is dynamic), which has clear advantages for modeling and understanding of the basic features. One is 
therefore limited to systems in the vicinity of the vanishing baryon, strangeness, or charge chemical potentials. Such systems, with nearly zero net charge 
densities, are formed in the mid-rapidity region of the ultra-relativistic heavy-ion collisions at the highest RHIC or the LHC energies.
Various susceptibilities, studied in this paper, bear significance to the investigation of the phase transitions and the critical point of QCD~\cite{Stephanov:1998dy,Stephanov:1999zu}, which have been actively studied on the lattice (see, e.g.,~\cite{Fodor:2004nz,Aoki:2006we,Bazavov:2011nk}),
importantly also in the presence of $B$~\cite{Ding:2023bft,Ding:2025jfz}. 
A notable associated phenomenon to mention here is the magnetic catalysis~\cite{Miransky:2002rp,Bali:2012zg,Mueller:2015fka}, proving nontrivial magnetic properties of the QCD vacuum itself. 

In a hot hadronic matter at vanishing chemical potentials, the QCD quark and gluon degrees of freedom dominate above the cross-over temperature $T_c \sim 155$~MeV. Below $T_c$, a remarkably good description of the thermodynamic properties of the medium is achieved within the Hadron Resonance Gas (HRG) model~\cite{Fermi:1950jd,Pomeranchuk:1951ey,Hagedorn:1965st,Koch:1986ud,Dashen:1969ep,Venugopalan:1992hy} (for the historical development and more references see, e.g., \cite{Florkowski:2010zz}). Its
most prominent success (at $B=0$) is the description of the hadron multiplicities measured in the ultra-relativistic nuclear 
collisions (for a review, see~\cite{Andronic:2005yp}). When amended with flow, HRG describes the features of the soft $p_T$ spectra as well~\cite{Broniowski:2001we,Florkowski:2010zz}. Importantly, the model also reproduces to expected accuracy various charge (baryon, strangeness, electric charge) susceptibilities (diagonal and non-diagonal) as functions of $T$ at $B=0$~\cite{Jeon:2000wg,Asakawa:2000wh,Borsanyi:2011sw,Bellwied:2015lba,Borsanyi:2018grb}, matching smoothly to the lattice QCD simulations up to $T_c$.

The susceptibilities from the lattice QCD at $B$ up to $0.8~{\rm GeV}^2$ and for $T=145$, $155$, and $165$~MeV 
were recently presented in~\cite{Ding:2025jfz}, extending the results of~\cite{Ding:2023bft}.
In these works, the comparisons to HRG were also made. The HRG analysis of~\cite{Ding:2023bft} was further developed in dedicated studies of~\cite{Marczenko:2024kko,Vovchenko:2024wbg}.
In all these papers, following the work of~\cite{Endrodi:2013cs,Bali:2014kia,Bali:2020bcn}, the effects of anomalous magnetic moments of hadrons were 
neglected.
In particular, for spin-1/2 hadrons the Lande $g$ factors were taken as $g=2Q$, where $Q$ is the electric charge. This calls for improvement, as one expects large effects in the baryon susceptibilities from the nucleons, where the $g$ factor of the proton $g_p\simeq  2\times 2.79\simeq 5.6$ (a very large value compared to 2) and for the neutron $g_n\simeq -2\times 1.91 \simeq - 3.8$ (the neutron's $B$-dependent contribution was neglected in the above-mentioned HRG studies). 

In fact, the anomalous magnetic moments of hadrons were investigated much earlier within HRG in the context of susceptibilities by
Bhattacharyya, Ghosh, Ray, and Samanta~\cite{Bhattacharyya:2015pra}, prior to the availability of the lattice QCD data.
These authors apparently used the formula for the spin-$\tfrac{1}{2}$ particles with anomalous magnetic moments~\cite{Tsai:1971zma}, but details were not 
given.

In this paper we return to the issue of the diagonal and non-diagonal susceptibilities in HRG, investigating the role of 
the anomalous magnetic moments of hadrons and comparing the results to the now available lattice QCD data. 
We start by carefully discussing (Sec.~\ref{s:hrgmag}) the inclusion of anomalous magnetic moments, where a rigorous theoretical treatment is available 
only for elementary spin-$\tfrac{1}{2}$ states~\cite{Tsai:1971zma}, whereas the treatment of higher spin hadrons must be approximate.
In this context, we list the magnetic properties of the lowest mass hadrons, exhibiting, {\it i.a.}, a large departure of the $g$-factors of the octet baryons 
from the non-anomalous $2Q$ value. 
We also discuss till what value of $B$ the model is applicable.
Section~\ref{s:hrg} contains a glossary of the used standard thermodynamics 
formulas and the details of the version of HRG used in this work.
In Sec.~\ref{s:magmom} we present a simple estimate of the role of the anomalous magnetic moments in susceptibilities, and pass to their detailed 
anatomy in Sec.~\ref{s:anatomy}.  Our results are presented in Sec.~\ref{s:res}, first for $B=0$, to check how HRG with its choice of the hadron states works in this ``conventional'' case, and then passing to the non-zero $B$ case. 
As expected, large effects of the anomalous moments are found compared to the $g=2Q$ assumption, leading to improvements in the comparison to the lattice data~\cite{Ding:2023bft,Ding:2025jfz} in the studied domain of $B\le 0.15~{\rm GeV}^2$. For completeness we also discuss the effect of anomalous magnetic moment on hadron multiplicities in Sec.~\ref{s:mult}. 

Throughout this paper we work in the natural units $e=\hbar=c=k_B=1$.

\section{Energy spectra of hadrons in a~uniform magnetic field \label{s:hrgmag}}

In the presence of a constant uniform magnetic field $B$ (assumed to be aligned with the $z$-axis), two competing effects appear: The Landau diamagnetism~\cite{Landau:1980mil} from the motion of a charged particle of mass $M$ and charge $Q$, and the paramagnetism due to non-zero intrinsic magnetic moments of hadrons with spin $s>0$ (present for both charged and neutral baryons, and charged mesons). Non-relativistically, 
the Landau levels appearing via the minimal substitution in the non-relativistic kinetic term $p^2/(2M)$ have the form~\cite{Landau:1980mil} 
\begin{eqnarray}
\epsilon_{l,p_z} = M+\frac{p_z^2}{2M} + \frac{B |Q|}{2M}(2l+1), \label{eq:Lnr}
\end{eqnarray}
where $l=0,1,\dots$ labels the level.
A paramagnetic interaction adds the term 
\begin{eqnarray}
\epsilon_p = - \vec{\mu} \cdot \vec{B}
\end{eqnarray}
to each Landau level, with $\vec{\mu}$ denoting the magnetic moment and $\vec{B}$ the magnetic field. Further,
\begin{eqnarray}
\mu = \mu_M g s_z, \;\; \mu_M=\frac{1}{2M}, \label{eq:Lande0}
\end{eqnarray}
where $\mu_M$ is the {\it natural magneton} (reflecting the mass of a given particle), $s_z=-s, \dots, s$ is the spin projection, and $g$ denotes the 
Lande $g$-factor for the hadron. Since the experimental values of $\mu$ are by convention defined for the highest value of spin, $s_z=s$, 
one has 
\begin{eqnarray}
\mu_{s_z=s}\equiv \mu_{\rm exp} = \mu_M g s. \label{eq:Lande1}
\end{eqnarray}
Typically, the experimental values for all states are quoted in nuclear magnetons $\mu_N=1/(2 m_p)$, hence from Eqs.~(\ref{eq:Lande0},\ref{eq:Lande1})
\begin{eqnarray}
g_{\rm exp} = \frac{\mu_{\rm exp}}{s \mu_M} = \frac{\mu_{\rm exp}}{s \mu_N} \frac{M}{m_p}. \label{eq:Lande2}
\end{eqnarray}

The values of the magnetic moments and the corresponding $g$-factors for the lowest 
mass states used in this paper are collected in Table~\ref{tab:hadrons} (other states used are taken with $g=2Q$).
We recall that because the magnetic moment operator is odd under the charge conjugation, the magnetic moments of particles and antiparticles are 
equal and opposite. In particular, the magnetic moments of neutral mesons vanish.

To recapitulate the above part, in the non-relativistic treatment the formula for the energy level of a charged hadron of {\it any spin} is
\begin{eqnarray}
E_{l,p_z}^{\rm nr} = M+\frac{p_z^2}{2M}+\frac{B}{2M}\left[|Q|(2l+1)- g s_z \right], \label{eq:enernr}
\label{eq:emagnonrel}
\end{eqnarray}
whereas for a neutral hadron 
\begin{eqnarray}
E_{p}^{\rm nr} = M+\frac{p^2}{2M}-\frac{B}{2M} g s_z.
\label{eq:emagnonrel0}
\end{eqnarray}
However, in ultra-relativistic nuclear collisions the large temperatures imply large momenta of hadrons, typically of a few hundred MeV, hence for a successful 
phenomenology of HRG it is required to use the relativistic kinematics. Below, we discuss one-by-one the subsequent spin states, keeping in mind that 
the full hadron list contains states up to as large a value as $s=\tfrac{11}{2}$, with the higher $s$ states suppressed by mass, but enhanced by their spin degeneracy factor $(2s+1)$.

\begin{table}
\caption{Magnetic moments and the corresponding  $g$-factors from Eq.~(\ref{eq:Lande2}) for the lowest mass hadrons, collected from various experimental, lattice QCD, and theoretical estimates. In the reference column the corresponding sources are listed. In case of multiple values of $\mu$ given, the value used in our calculation is indicated in bold. While the value of the anomalous magnetic moments for resonances are in general complex, we quote and use the real values. \\ \label{tab:hadrons}}
\begin{tabular}{|c|c|c|c|}
\hline
   Hadron species &  $\mu/\mu_N$  & $g$ (Eq.~\ref{eq:Lande2}) & Reference \\ \hline
   $\rho^+(775)$ & {\bf 1.94}(1)  & 1.60(1) & LQCD~\cite{QCDSF:2008tjq}\\
   & 2.21  & 1.82 &  $\chi$PT~\cite{Djukanovic:2013mka} \\ 
   
   & 2.37 & 1.96 & QM~\cite{Badalian:2012ft} \\  \hline
   
   $K^{*+}(892)$ & {\bf 2.4}(2)& 2.3(2) & LQCD~\cite{Luschevskaya:2018sbp} \\ 
   & 2.19 & 2.08 &QM~\cite{Badalian:2012ft}\\ 
    \hline
   $K^{*0}(896)$ &  -0.183 & -0.175 & QM~\cite{Badalian:2012ft}\\ \hline
   $p(938)$ &  2.793 &  5.586 &PDG~\cite{ParticleDataGroup:2024cfk} \\ \hline
   $n(939)$ &  -1.913 & -3.831 &PDG~\cite{ParticleDataGroup:2024cfk}  \\ \hline
   $\Lambda^0(1115)$  & -0.613(4) & -1.458(9)& PDG~\cite{ParticleDataGroup:2024cfk} \\ \hline
   $\Sigma^+(1189)$ & 2.458 (10) & 6.232(25)  &PDG~\cite{ParticleDataGroup:2024cfk}\\ \hline
   $\Sigma^0(1192)$ &{\bf 0.65} & 1.65 & $\chi$PT~\cite{Meissner:1997hn} \\
   & 0.791 & 2.011& QM~\cite{ParticleDataGroup:2024cfk}\\ 
    \hline
   $\Sigma^-(1197)$ &  -1.160(25) & -2.96(63) &PDG~\cite{ParticleDataGroup:2024cfk}\\ \hline
   $a_1^+(1230)$ & {\bf 1.7}(2) & 2.2(2)&LQCD~\cite{Lee:2008qf}\\
   & 1.44& 1.89 & QM~\cite{Badalian:2012ft}\\ 
    \hline 
   $\Delta^{++}(1232)$ & {\bf 3.7-7.5} & 3.23-6.56  & PDG~\cite{ParticleDataGroup:2024cfk} \\
   & {\bf 6.14}(51) & 5.37(45) & Lopez et.al~\cite{LopezCastro:2000cv} \\
    & {\bf 5.24}(18) & 4.58(16) & LQCD~\cite{Lee:2005ds} \\ 
   & {\bf 4.97(89)} & 4.34(78) & $\chi$PT~\cite{Li:2016ezv} \\  \hline
   $\Delta^{+}(1232)$ & $2.7^{+1.0}_{-1.3}\!\pm\! 1.5 \!\pm \!3$ & $2.36^{+0.87}_{-1.14}$  & PDG~\cite{ParticleDataGroup:2024cfk} \\
   & {\bf 2.6(5)} & 2.27(4) & $\chi$PT~\cite{Li:2016ezv}\\ 
     \hline 
   $\Delta^0(1232)$ & 0.02(12) & 0.017(100) & $\chi$PT~\cite{Li:2016ezv}\\ \hline
   $\Delta^{-}(1232)$ & -2.48(32) & -2.17(28) & $\chi$PT~\cite{Li:2016ezv} \\ \hline
   $\Xi^-(1321)$ & -0.651(3) & -1.834(8) & PDG~\cite{ParticleDataGroup:2024cfk}\\ \hline
   $\Xi^0(1321)$ & -1.250(14) & -3.503(39) & PDG~\cite{ParticleDataGroup:2024cfk}\\ \hline
   $\Sigma^{*+}(1383)$ &{\bf 2.55}(26)& 2.50(25)& LQCD~\cite{Leinweber:1992hy} \\
   & 1.76(38) & 1.73(37) &$\chi$PT~\cite{Li:2016ezv} \\
     \hline
   $\Sigma^{*0}(1384)$ &{\bf 0.27}(5) & 0.26(5) & LQCD~\cite{Leinweber:1992hy} \\
   & -0.02(3) & -0.02(3) & $\chi$PT~\cite{Li:2016ezv}\\ 
    \hline
   $\Sigma^{*-}(1387)$ &{\bf -2.02}(18)& -1.99(18) & LQCD~\cite{Leinweber:1992hy} \\
   &  -1.85(38) & -1.82(37) &$\chi$PT~\cite{Li:2016ezv}\\
    \hline
   $N^+(1440)$ &  0.48 & 1.47 & $\chi$PT~\cite{Bauer:2012at} \\ \hline
   $N^0(1440)$ &  0.27 & 0.84 & $\chi$PT~\cite{Bauer:2012at} \\ \hline
   $\Xi^{*-}(1530)$ &{\bf-1.68}(12)& -1.82(13) & LQCD~\cite{Leinweber:1992hy} \\
   & -1.90(47) & -2.07(51) & $\chi$PT~\cite{Li:2016ezv} \\
    \hline
   $\Xi^{*0}(1530)$ &{\bf 0.46}(7)& 0.50(8) & LQCD~\cite{Leinweber:1992hy} \\
   & -0.42(13) & -0.46(14) & $\chi$PT~\cite{Li:2016ezv} \\
    \hline
   $\Omega^-(1672)$ & -2.02(5) & -2.40(6) & PDG~\cite{ParticleDataGroup:2024cfk}\\ \hline
\end{tabular}
\end{table}

The spin-0 states obviously have no magnetic moment, as follows from the Wigner-Eckart theorem. In this case the minimum coupling in the 
Klein-Gordon equation
yields for the charged mesons of charge $Q$
\begin{eqnarray}
\hspace{-4mm} \epsilon_{l,p_z}=\sqrt{M^2+p_z^2 + B |Q| (2l+1)},\; (s=0, {\rm ~charged}), \;\;
\end{eqnarray}
and obviously $\epsilon_{p}=\sqrt{M^2+p^2 }$, independent of $B$,  for the neutral mesons. 

For the spin-$\tfrac{1}{2}$ Dirac states (describing, i.a., the octet baryons) with the non-minimal (anomalous) coupling of the form $\mu_M \kappa \bar{\psi} \sigma_{\mu \nu} F^{\mu \nu}\psi$, the eigenvalue problem was first solved 
by Tsai and Yildiz~\cite{Tsai:1971zma}, with the result (we use the convention where $g=2(Q+\kappa)$)
\begin{eqnarray}
&&\hspace{-2mm} \epsilon_{l,p_z}=\sqrt{\left (\sqrt{M^2+ B (2l\!+\!1\!-\!2Q s_z)}-2\mu_M B\kappa s_z\right)^2+p_z^2} \nonumber \\ 
&&\hspace{4cm} (s=\tfrac{1}{2}, Q=\pm 1). 
\end{eqnarray}
For $Q=0$, an elementary calculation yields
\begin{eqnarray}
&&\hspace{-3mm} \epsilon_{p,p_z}=\sqrt{\left (\sqrt{M^2+ p^2-p_z^2}-2\mu_M B\kappa s_z\right)^2+p_z^2} \nonumber \\ 
&&\hspace{4cm} (s=\tfrac{1}{2}, Q=0). 
\end{eqnarray}

For the spin-1 states, the energy levels take a simple form only for the case  $g=2Q$, which in this case requires a non-minimal coupling in addition to the minimum substitution 
in the Proca field Lagrangian (cf. Eq.~(17) of~\cite{Tsai:1971zma} with their $\kappa=1$, and $g=1+\kappa$ for the assumed $Q=1$).
This is in agreement with the general conviction discussed in~\cite{Ferrara:1992yc} that at the tree level $g=2$ (or, in general $2Q$, when charges different from 1 are admitted)
for point-like particles of any spin $s$, rather than the earlier expectations of $g=Q/s$~\cite{Belinfante:1953zz}.
For the spin-1 mesons, as seen from Table~\ref{tab:hadrons}, 
the $g$-factor is indeed close to $2$. Thus we adopt  Eq.~(17) of~\cite{Tsai:1971zma}, supplementing it with a small correction 
accounting for
the departure of $g$ from the $2Q$ value in the non-relativistic form of Eq.~(\ref{eq:enernr}), namely
\begin{eqnarray}
&&\hspace{-10mm} \epsilon_{l,p_z}=\sqrt{M^2+p_z^2+ B [|Q|(2l\!+\!1\!)-\!2Q s_z]} \nonumber \\
&&\hspace{1cm} - B \mu_M (g-2Q) s_z, \;\;\;
(s=1, Q \neq 0). 
\end{eqnarray}

For the spin-$\tfrac{3}{2}$ Rarita-Schwinger states with $g=2Q$, it was argued in~\cite{dePaoli:2012eq} that the 
energy spectrum is given by $\epsilon_{l,p_z}=\sqrt{M^2+p_z^2+ B [|Q|(2l+1)-2Q s_z]}$. Extension to account for the anomalous case is 
difficult (see, e.g., the early work~\cite{Seetharaman:1975dq}). Fortunately, for the $\Delta(1232)$ states listed in Table~\ref{tab:hadrons}, $g$ is close to the $2Q$ value. To include a small departure of $g$ from this value, 
similarly to the spin-1 case above, we take
\begin{eqnarray}
&&\hspace{-8mm} \epsilon_{l,p_z}=\sqrt{M^2+p_z^2+B [|Q|(2l\!+\!1\!)-\!2Q s_z]} \nonumber \\
&&\hspace{1cm}- B \mu_M (g-2Q) s_z, \;\;\; (s=\tfrac{3}{2}, Q \neq 0). 
\end{eqnarray}
For the neutral case we use, correspondingly,
\begin{eqnarray}
\hspace{-2mm} \epsilon_{p}=\sqrt{M^2+p^2} - B \mu_M g s_z, \;(s=\tfrac{3}{2}, Q = 0). 
\end{eqnarray}

For hadrons of spins higher than $\tfrac{3}{2}$, which are heavier from the lower spin states (the lightest state is $a_2(1320))$, we use, 
with the lack of detailed knowledge and guided by the lower-spin cases, the expressions
\begin{eqnarray}
&& \hspace{-10mm} \epsilon_{l,p_z}=\sqrt{M^2+p_z^2 + B |Q|(2l+1)} - B \mu_M g s_z \nonumber \\
&& \hspace{4cm} (s>\tfrac{3}{2}, Q\neq 0),
\end{eqnarray}
and
\begin{eqnarray}
\hspace{-2mm} \epsilon_{p}=\sqrt{M^2+p^2} - B \mu_M g s_z, \;\; (s>\tfrac{3}{2}, Q=0),
\end{eqnarray}
assuming $g=2Q$. 

The policy of including the hadron states in HRG in the uniform magnetic field according to the above formulas is in our view the best phenomenological 
approach that we have at hand. It is approximate in several aspects: First, the hadrons are not point-like, hence their structure such as the mass may 
additionally depend on $B$. Second, the $g$-factors for finite-width resonances are complex numbers, so the formulas are valid for the zero-width case. 
Also, the position of a resonance and its width in general can depend on $B$. Finally, at very large values of $B$ the square of the energies of the low Landau levels may turn negative, exhibiting instability~\cite{Endrodi:2013cs} and indicating the limits of applicability of the approach.
For instance, the lowest Landau level for $\Delta^{++}$ with $s_z=\tfrac{3}{2}$ turns imaginary at $B \simeq 0.3~{\rm GeV}^2$, providing 
a rough upper limit for applicability of the approach with structureless hadrons. 

Taking a measure of a basic excitation energy in the nucleon as $m_\Delta -m_N$ and requesting that the magnetic interaction in the proton be smaller, 
\begin{eqnarray}
\mu_p B \lesssim m_\Delta-m_p,
\end{eqnarray}
yields the applicability condition 
\begin{eqnarray}
B \lesssim 0.2~{\rm GeV}^2, \label{eq:upperB}
\end{eqnarray}
more stringent than the one following from the instability, discussed above.

Concerning the magnetic effects on the mass to the second order in low $B$, one can make an estimate for the nucleon making use of the known values of 
the magnetic polarizabilities. The relevant formula here is 
\begin{eqnarray}
\Delta M = -\frac{1}{2}\beta B^2,
\end{eqnarray}
where $\beta$ denotes the magnetic polarizability. For the proton and neutron we have $\beta_p=2.5(4)\times 10^{-4}~{\rm fm}^3$ 
and $\beta_n=3.7(1.2)\times 10^{-4}~{\rm fm}^3$, which at our top $B$ value of $0.15~{\rm GeV}^2$ leads to mass shifts $\sim 0.5$~MeV, hence 
completely negligible at temperatures of the order of $T_c$. However, the conclusion is no longer valid at stronger $B$ fields, where higher-order effect may 
take over. In fact, estimates made within the Skyrme model show that in strong magnetic fields the deformation effects lead to a substantial increase of the nucleon and $\Delta$ masses~\cite{Yakhshiev:2019gvb,Chen:2023gws}. Such effects, not attempted to be incorporated here, would quench the susceptibilities at larger values of $B$, as well as would help with the instability problem mentioned above.

\section{HRG in a uniform magnetic field \label{s:hrg}}

In this section we provide the standard expressions (see, {\it e.g.}, \cite{pathriaStatisticalMechanics2011}) for the thermodynamics of HRG, needed in our work. The thermal partial pressure of the individual species of HRG can be written as~\cite{Endrodi:2013cs,Marczenko:2024kko,Vovchenko:2024wbg} 
\begin{eqnarray}
\hspace{-8mm} P= -\eta T (2s+1) \int \frac{d^3p}{(2\pi)^3} \ln[1-\eta f(E,T,\mu)],
\label{eq:pressure}
\end{eqnarray}
where $f(E,T,\mu)$ is the distribution function 
\begin{eqnarray}
f(E,T,\mu) = \frac{1}{\exp(\frac{E-\mu}{T})+ \eta},
\label{eq:distfunc}
\end{eqnarray}
with $\mu = \mu_B B + \mu_S S +\mu_Q Q$ representing the total chemical potential of a single hadron,
 and $\eta=+1 \ \text{or} -1$, depending on whether we deal with a fermion or a boson. In the presence of a magnetic field, for charged hadrons the phase space integral is written as the sum over the Landau levels and the spin states~\cite{Andersen:2014xxa}
\begin{eqnarray}
(2s+1) \int \frac{d^3p}{(2\pi)^3} \longrightarrow \frac{B|Q|}{2\pi^2} \sum_{l=0}^\infty \sum_{s_z=-s}^s \int_0^\infty dp_z,
\label{eq:phspace}
\end{eqnarray}
while  for the neutral hadrons 
\begin{eqnarray}
(2s+1) \int \frac{d^3p}{(2\pi)^3} \longrightarrow \frac{1}{2\pi^2} \sum_{s_z=-s}^s \int_0^\infty p^2 dp.
\label{eq:phspaceneu}
\end{eqnarray}
Then for the charged particles, the partial pressure $P$ takes the form
\begin{eqnarray}
P= -\eta T \frac{B|Q|}{2\pi^2} \sum_{l=0}^\infty \sum_{s_z=-s}^s \int_0^\infty dp_z \ln[1-\eta f],
\label{eq:pressurefull}
\end{eqnarray}
whereas for neutral particles 
\begin{eqnarray}
P= -\eta T \frac{1}{2\pi^2} \sum_{s_z=-s}^s \int_0^\infty p^2 dp \ln[1-\eta f].
\label{eq:pressureneu}
\end{eqnarray}

The leading order conserved charge (dimensionless) susceptibilities are defined as
\begin{eqnarray}
\chi_{Q_1Q_2}= \frac{\partial^2 (P/T^4)}{\partial(\mu_{Q_1}/T)\partial(\mu_{Q_2}/T)} \Bigg |_{T} \ ,
\label{eq:suscdef}
\end{eqnarray}
where $Q_1,Q_2 = \{B,S,Q\}$ indicate the type of the conserved charge. Using Eq.~(\ref{eq:pressurefull}) and (\ref{eq:pressureneu}) in Eq.~(\ref{eq:suscdef}) immediately yields
\begin{eqnarray}
\hspace{-7mm} \chi_{Q_1Q_2}= \frac{Q_1Q_2B|Q|}{2\pi^2T^3} \sum_{l=0}^\infty \sum_{s_z=-s}^s \int_0^\infty dp_z f(1-\eta f), \label{eq:su1}
\label{eq:suscfull}
\end{eqnarray}
for the charged particles and
\begin{eqnarray}
 \chi_{Q_1Q_2}= \frac{Q_1Q_2}{2\pi^2T^3} \sum_{s_z=-s}^s \int_0^\infty p^2 dp f(1-\eta f)  \label{eq:su2}
\label{eq:suscfull0}
\end{eqnarray}
for the neutral case. 

A remark regarding how we handle numerically the calculation of the susceptibilities of Eqs.~(\ref{eq:suscfull},\ref{eq:suscfull0}) is in order. At a given $T$ and $B$, a certain number of Landau levels (up to $l_{max}\sim 1/B$) effectively saturates the infinite sum. We thus sum as many levels as needed to achieve convergence, which for any $B>0$ is straightforward. For $B=0$ strictly, one of course uses the formulas without the magnetic field following Eqs.~(\ref{eq:pressure},\ref{eq:distfunc}). 

For a critical review of the HRG model and its modifications, the reader is referred to, {\it e.g.},~\cite{Bollweg:2021vqf}. 
Despite the caveats discussed there, the success of the HRG model in the description of both the hadron multiplicities in ultra-relativistic nuclear collisions, as well as the properties of the hot medium in its validity range of $100~{\rm MeV} \lesssim T \lesssim 150~{\rm MeV}$, places it as an established phenomenological tool or even as a benchmark in certain comparisons to data.

We use the hadron list QMHRG2020~\cite{Bollweg:2021vqf,Ding:2023bft}, which contains an extended list of resonance states based on the quark model (including the so-called missing states)
compared to the Particle Data Group listings~\cite{ParticleDataGroup:2024cfk}. The PDG list, which has less states, fails to reproduce with an expected accuracy the lattice data at zero magnetic field~\cite{Bollweg:2021vqf}, hence QMHRG2020 is preferable.

The magnetic moments of the lowest mass hadrons are collected in Table~\ref{tab:hadrons}. We note the large anomalous magnetic moments of the octet baryons ($N$, $\Lambda$, $\Sigma$, and $\Xi$), and the approximate proximity to the $g=2Q$ values for the decuplet baryons 
($\Delta$, $\Sigma^\ast$, $\Xi^\ast$, and $\Omega$). We also remark that for the Roper state, $N^+(1440)$, 
the $g$-factor is somewhat below 2. For the $\rho^+$,  $K^{\ast +}$, and $a_1^+$ mesons, the corresponding  $g$-factors are close to 2. 
These results support the view that except for the octet baryons, which are spin-$\tfrac{1}{2}$ and for which we have a formula with an exact treatment of the anomalous magnetic moments, the remaining hadrons abide to the $g\simeq 2Q$ case, which can be treated to a good accuracy with the formulas 
collected in Sec.~\ref{s:hrgmag}.

\section{Relevance of physical magnetic moments of hadrons \label{s:magmom}}

In the non-relativistic treatment of Eq.~(\ref{eq:emagnonrel}), which we apply in this Section for the simplicity of the discussion, and neglecting, for clarity, the effects of statistics
in final expressions ({\it i.e.}, setting $\eta^2=1$ and $\eta=0$ after evaluation of the derivatives with respect to the chemical potentials), one easily gets the formula for the generic susceptibility,
\begin{eqnarray}
\hspace{-7mm} \chi_{Q_1 Q_2}= \frac{Q_1 Q_2 B |Q| e^{-\frac{M}{T}} \sqrt{M T} \sinh \left(\frac{B g  (2 s+1)}{4 M T}\right)}{4 \sqrt{2} \pi
   ^{3/2} T^3 \sinh\left(\frac{B |Q|}{2 M T}\right) \sinh\left(\frac{B g }{4 M T}\right)}.  \label{eq:chis}
\end{eqnarray} 
Here we write it for the electric charge $Q\neq 0$, but the limit of $Q\to 0$ gives 
the same expression for the neutral particles as well. Several interesting features can be inferred from this formula.  

For $s=0$ states, such as the pion or kaon, Eq.~(\ref{eq:chis}) behaves, as a function of $B$, as $B |Q|/\sinh(B|Q|/2MT)$, which is monotonically {\it decreasing} to zero. On the other hand, for $s > 0$, the last factors in the numerator and denominator of Eq.~(\ref{eq:chis}) yield an increase. The change of the behavior from decreasing to increasing occurs 
around $s \sim |Q/g|$. More precisely, from the expansions at low $B$,
\begin{eqnarray}
&& \chi_{Q_1 Q_2}=\frac{Q_1 Q_2 M^{3/2}  (2 s+1) e^{-\frac{M}{T}}}{2 \sqrt{2} \pi ^{3/2} T^{3/2}} \times \label{eq:chise}\\
&& \hspace{12mm}\left [1+ \frac{B^2 \left(g^2 s (s+1)-Q^2\right)}{24 M^2 T^2} \right ] 
+{\cal O} \left ( \frac{B^4}{M^2 T^2} \right ), \nonumber
\end{eqnarray}
we conclude that the function is concave when $g^2 s (s+1)<Q^2$, and convex otherwise. 
The above conclusions on the behavior of $\chi_{Q_1 Q_2}(B)$ remain qualitatively valid for the relativistic case and 
with the inclusion of the Bose-Einstein or Fermi-Dirac statistics.

There is one more important observation we want to underscore. Let us introduce
\begin{eqnarray}
\Delta \chi_{Q_1 Q_2}(B)= \chi_{Q_1 Q_2}(B) - \chi_{Q_1 Q_2}(0), 
\label{delchi}
\end{eqnarray}
which represents the difference between susceptibility at non-zero $B$ and $B=0$. Due to the subtraction of the \mbox{$B=0$} contribution, such a quantity shows the influence of a physical $g$ more prominently. As long as $B$ is sufficiently small (i.e., keeping the terms up to ${\cal O}(B^2)$), from  Eq.~(\ref{eq:chise}) we can see that for a given hadron the departure of 
$\chi_{Q_1 Q_2}$ from its value at $B=0$ is controlled by the pre-factor $g^2 s (s+1)-Q^2$. With the prescription $g=2Q$, this pre-factor becomes $[4s(s+1)-1]Q^2$. 

The change between using the physical values of $g$ and $g=2Q$ can thus be quantified through the ratio
\begin{eqnarray}
R_{Q_1Q_2}(B)=
\frac{\Delta \chi_{Q_1 Q_2}(B)_{g {\rm ~phys.} }}{\Delta \chi_{Q_1 Q_2}(B)_{g =2Q}}, \label{eq:ratio}
\end{eqnarray}
 which is independent of $B$ at low $B$. For a single state, according to the previous expressions, it is (non-relativistically and without statistics) 
 \begin{eqnarray}
R_{Q_1Q_2}(B)\simeq\frac{[\frac{g^2}{Q^2}s (s+1)-1]}{[4s(s+1)-1]}, 
\label{eq:ratios}
\end{eqnarray}
For the case of $\chi_{BB}$, the ratio $R_{BB}$ is dominated 
by the nucleons and the $\Delta(1232)$ resonances, which are the lightest baryons (cf. the following discussion of Fig.~\ref{fig:histchiBB}). For the proton $g_p\simeq 5.6$, and Eq.~(\ref{eq:ratios}) for this state yields an enhancement as large as $R \simeq 10$. It is to be mentioned that with the use of proper statistics and relativistic formula as discussed in Sec.~\ref{s:hrgmag}, this factor is even larger, $R\sim 14$.
For the neutron $g_n\simeq-3.8$, and the relative contribution of $\chi_{BB}$ of the neutron to the proton is about 0.5, hence substantial. We recall that since the neutron has no charge, its corresponding contribution is not included in the $g=2Q$ prescription at all. 
For $\Delta^{++}$, using the value of $g$ from Table~\ref{tab:hadrons} compared to $2Q=4$, leads to an enhancement by about 20\%, hence moderate. 

\section{Anatomy of the susceptibilities \label{s:anatomy}}

\begin{figure}[tb]
\begin{center}
\includegraphics[width=\linewidth]{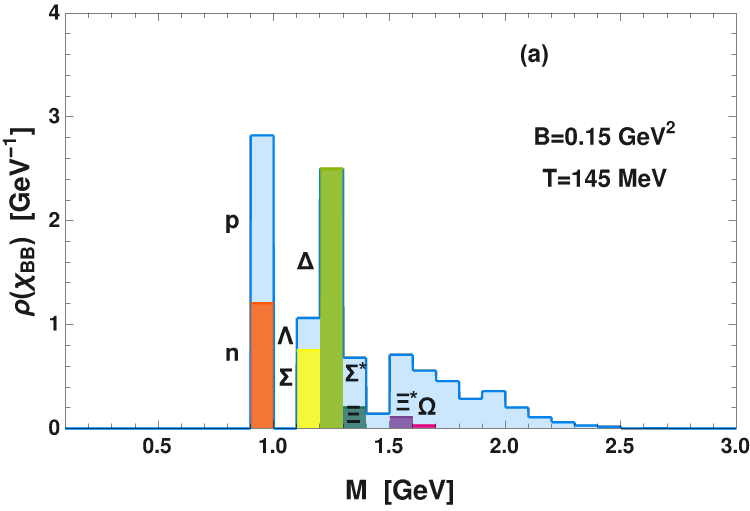}\\
\includegraphics[width=\linewidth]{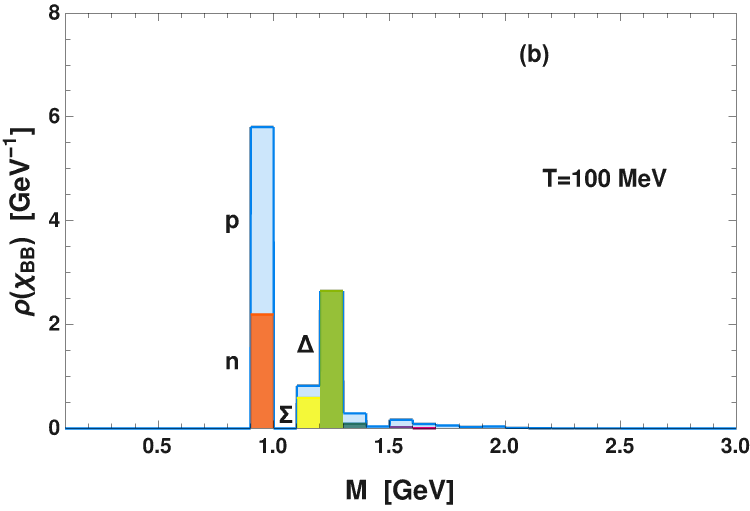} 
\end{center}
\caption{Normalized histogram density for the baryon number susceptibility $\chi_{BB}$ in HRG, plotted as a function of the hadron 
mass $M$ at $T=145~{\rm MeV}$ (a) and $T=100~{\rm MeV}$ (b), and at magnetic field strength $B=0.15~{\rm GeV}^2$. The HRG contribution in each bin is shown in light blue color, while the individual contributions from baryon octet and decuplet states are shown in different colors
and are labeled accordingly. 
\label{fig:histchiBB}}
\end{figure}

\begin{figure}[tb]
\begin{center}
\includegraphics[width=\linewidth]{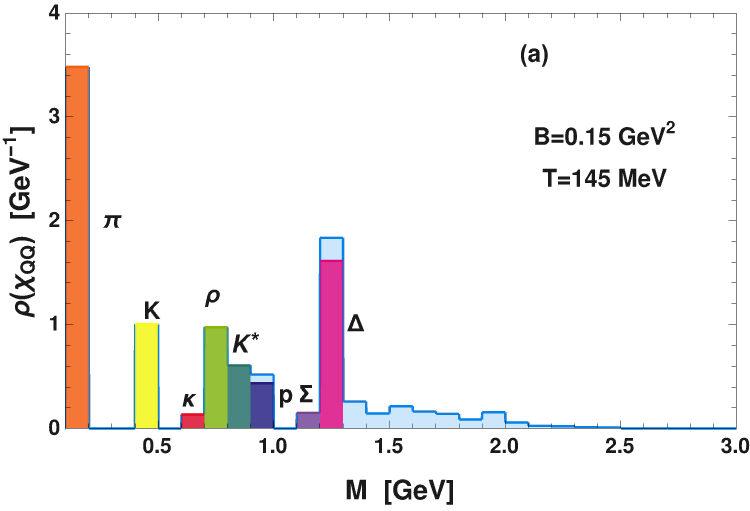}\\
\includegraphics[width=\linewidth]{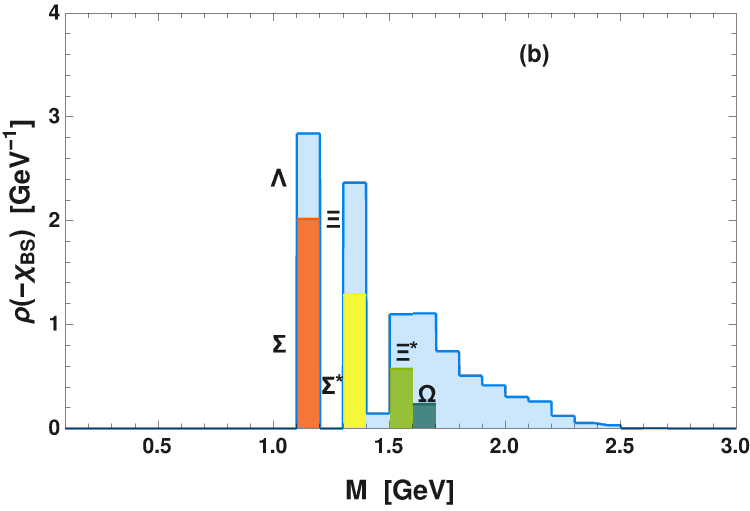} 
\end{center}
\caption{Same as Fig.~\ref{fig:histchiBB}, but for the electric charge susceptibility $\chi_{QQ}$ and the non-diagonal 
baryon-strangeness susceptibility $\chi_{BS}$ at $T=145~{\rm MeV}$.
\label{fig:histchiQQ}}
\end{figure} 
 
For a better perspective and to highlight the key motivation behind this paper, 
we look closer at the anatomy of $\chi_{Q_1Q_2}$, breaking it into the contributions of the
subsequent states, binned with width of 100~MeV in the mass $M$. 
The resulting histogram densities (distributions in $M$), denoted by $\rho(\chi_{Q_1Q_2})$, are 
normalized as $\int dM\, \rho(\chi_{Q_1Q_2})=1$. 
Figure~\ref{fig:histchiBB} shows $\rho(\chi_{BB})$ at $B=0.15$~GeV$^2$ for two different temperatures, $T=145$ MeV and 100 MeV, where the contributions from individual octet and decuplet baryons are shown in different colors. At $T=145$ MeV (panel~(a)), which is a value close to the 
cross-over phase transition, 
the dominant contribution to $\chi_{BB}$ comes from the nucleons, followed by the $\Delta$ resonance states, amounting jointly to 
about $\sim 50 \%$ of the total, while the rest $\sim 50 \%$ can be attributed to the other states, including a $\sim 10 \%$ contributions from the $\Lambda$ and $\Sigma$ baryons. 
It is instructive to note that with the use of the physical values of $g$ (as given in Table~\ref{tab:hadrons}), compared to the $g=2Q$ prescription, the proton 
contribution to $\chi_{BB}(T)$ is enhanced by a factor of $\sim 2$, and for the neutron, by more than $50 \%$, which is very substantial. The neutron takes about $\sim 40\%$ of the total nucleon contribution. For the $\Delta$, the analogous enhancement is around $\sim 20 \%$. 

As the temperature decreases, the higher mass states get suppressed with the thermal factors, hence the nucleons largely dominate. As seen from panel (b), for  $T=100$ MeV the nucleons alone contribute almost $\sim 60 \%$  of the total, the $\Delta$ resonances yield
$\sim 25 \%$ of the total, while the rest is essentially saturated with the remaining octet states. 

Similar histograms are presented in Fig.~\ref{fig:histchiQQ} for the electric charge susceptibility $\rho(\chi_{QQ})$ in panel (a) and for the non-diagonal 
baryon-strangeness susceptibility $-\chi_{BS}$ (note the inverted sign) in panel (b), at $T=145$ MeV for both cases. For $\chi_{QQ}$, the dominant contribution, naturally, comes 
from the  pions ($\sim 35 \%$), followed by the $\Delta$ states ($\sim 18 \%$). Together, $\pi$, $\Delta$, $\rho$ and $K$ yield almost $75 \%$ of the total.
The large contribution of the $\Delta$, despite the larger mass, comes from the electric charge of $\Delta^{++}$ and large spin and isospin degeneracy. As the temperature goes down (not shown), the pions dominate more and more, taking $\sim 70 \%$  at $T=100$ MeV. 

On the other hand, for $\chi_{BS}$ of panel (b), the dominant contribution comes from $\Sigma$ and $\Sigma^\ast$, taking more than $30\%$ of the total, followed by the $\Xi$ and $\Lambda$ baryons. As expected, in the anatomy of the strangeness susceptibility, $\chi_{SS}$ (not shown), the kaons provide the dominant contribution.    

Returning to the distributions of the susceptibilities  $\chi_{BB}$, $\chi_{QQ}$, and also $\chi_{BQ}$ (not shown), we note that the contribution of the $\Delta$ resonance to these quantities is large. At the same time, the accuracy of the  determination of magnetic moments of the decuplet states is not precise, hence $\Delta$ brings in a model uncertainty to the HRG predictions.

\section{Comparison to the lattice QCD data \label{s:res}}

\subsection{Susceptibilities at B=0 \label{s:suscB0}}

\begin{figure*}[h]
\begin{center}
\includegraphics[height=5 cm]{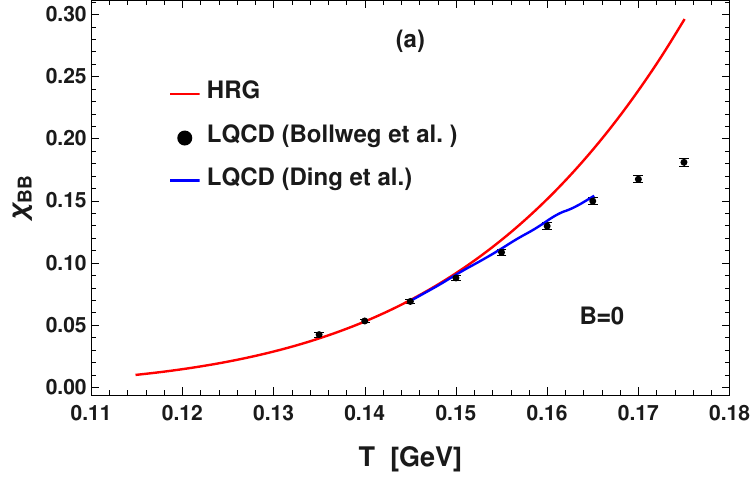}~~~~~\includegraphics[height=5 cm]{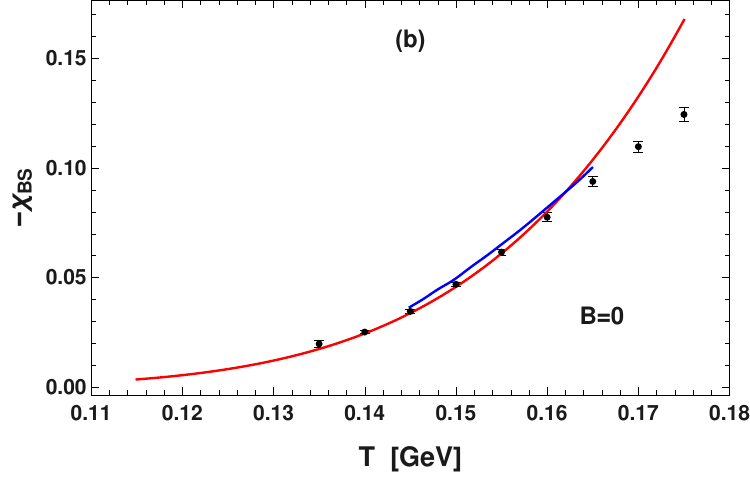} \\
\includegraphics[height=5 cm]{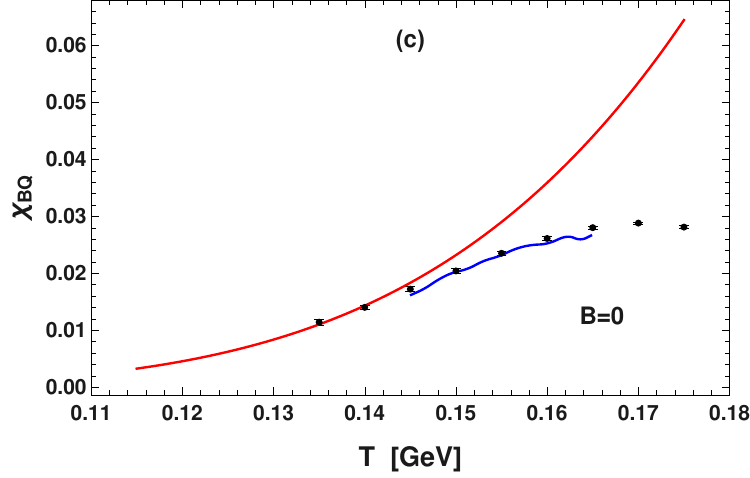}~~~~~\includegraphics[height=5 cm]{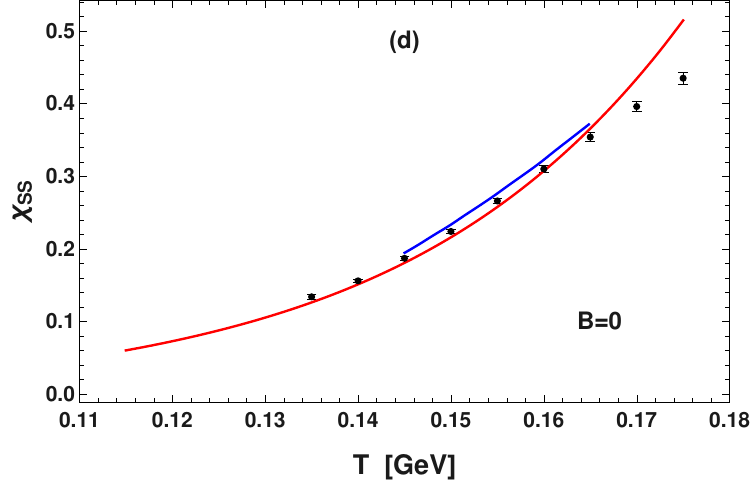} \\
\includegraphics[height=5 cm]{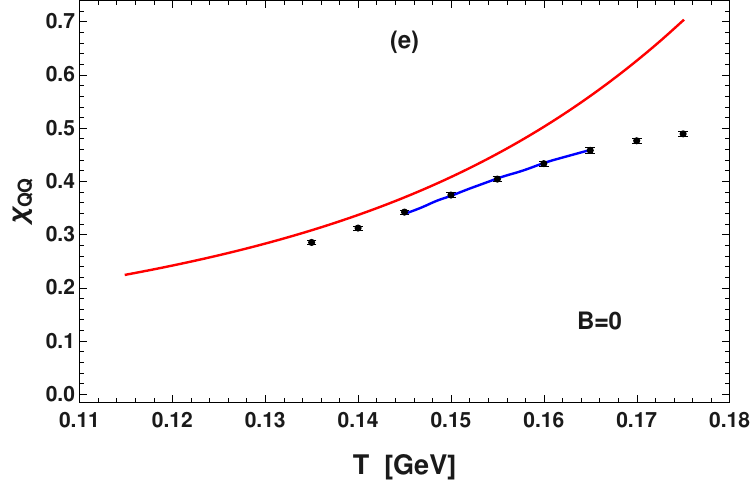}
~~~~~\includegraphics[height=5 cm]{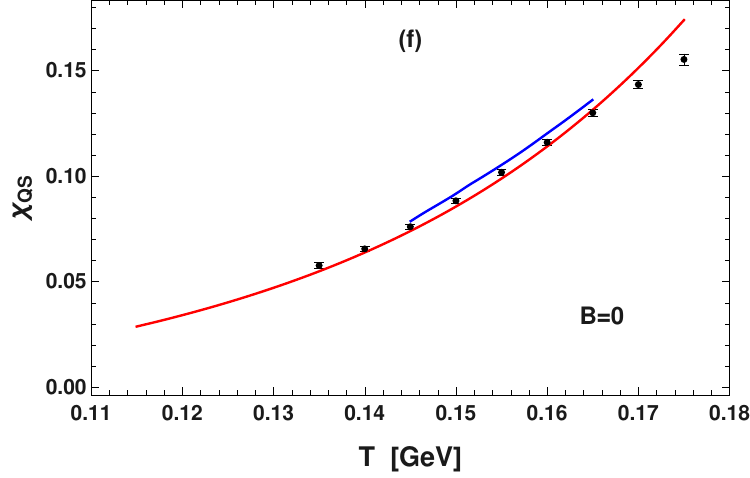}\\
\end{center}
\caption{Diagonal and non-diagonal conserved charge susceptibilities at zero magnetic field, plotted as functions of the temperature $T$. 
The red solid line represents our HRG model calculation, the black symbols denote the lattice data from~\cite{Bollweg:2021vqf}, 
and the blue curve represents the lattice data digitized from~\cite{Ding:2025jfz}. 
\label{fig:susczeroB}}
\end{figure*} 

We first study the susceptibilities in the absence of the magnetic field, i.e., for $B=0$, in order to 
see how HRG works in this basic case, before switching on the $B>0$ effects. Such a study was already done in~\cite{Bollweg:2021vqf}, 
but we independently repeat it here for completeness.
Figure~\ref{fig:susczeroB} shows the leading order diagonal and non-diagonal conserved charge susceptibilities as functions of $T$. The HRG model calculations (red lines) are compared with two sets of the lattice QCD data, one from~\cite{Bollweg:2021vqf} and the other one digitized from~\cite{Ding:2025jfz}. Both sets of data agree with HRG in the fiducial range of $T\lesssim 150$~MeV
within a few percent. It should be noted that for $-\chi_{BS}$, $\chi_{QS}$ and $\chi_{SS}$, HRG exhibits a somewhat better agreement with 
the data from~\cite{Bollweg:2021vqf} than from~\cite{Ding:2025jfz}. Beyond $\sim 155$~MeV there is an 
anticipated departure of the the model results from the data, which bend over to the quark-gluon plasma (QGP) phase results. The quality of the agreement between HRG and lattice data is somewhat worse in case of $\chi_{QQ}$, where the HRG calculation overestimates the lattice results by about 8 \%. This is an expected behavior of $\chi_{QQ}$ at $B=0$, also seen in past studies~\cite{Bollweg:2021vqf,Ding:2023bft,Vovchenko:2024wbg,Marczenko:2024kko}, and not a primary focus of this paper. However this discrepancy at $B=0$ is propagated to $B>0$ case, as we discuss in the next section.

\subsection{Susceptibilities at nonzero B \label{s:suscB}}

\begin{figure*}[h]
\begin{center}
\includegraphics[height=5 cm]{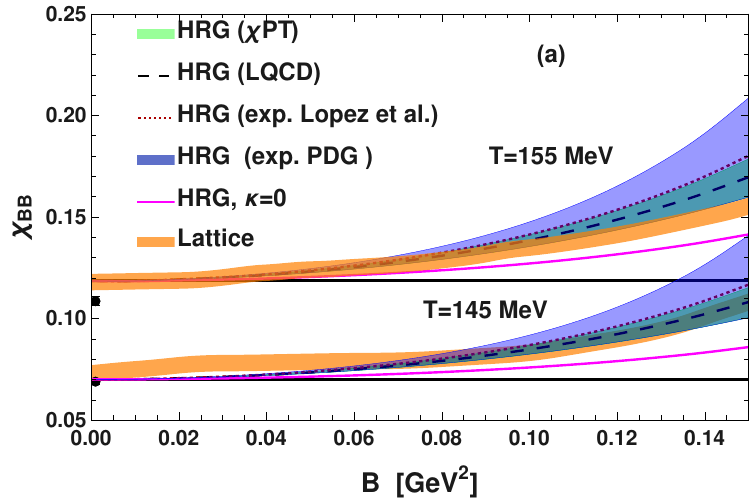}~~~\includegraphics[height=5 cm]{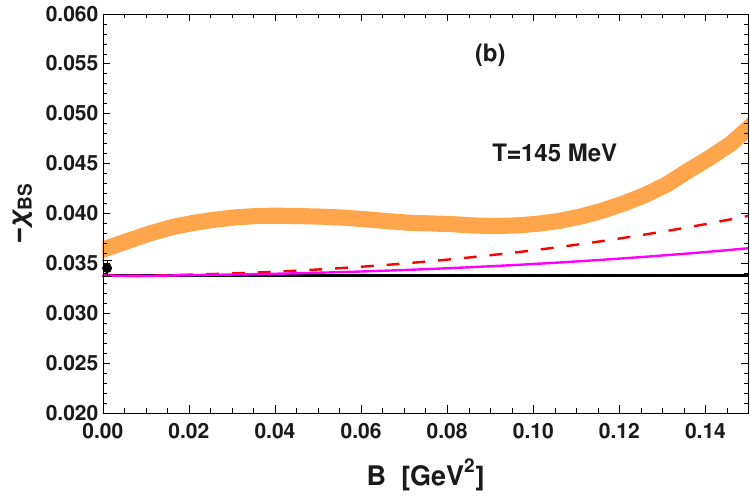} \\
\includegraphics[height=5 cm]{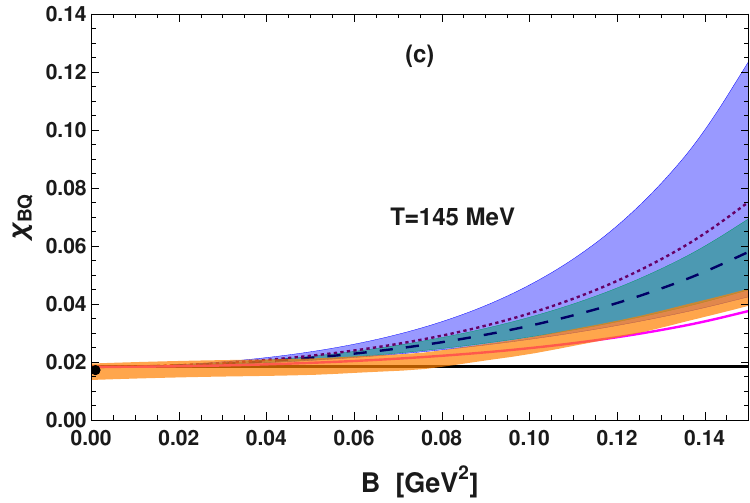}~~~\includegraphics[height=5 cm]{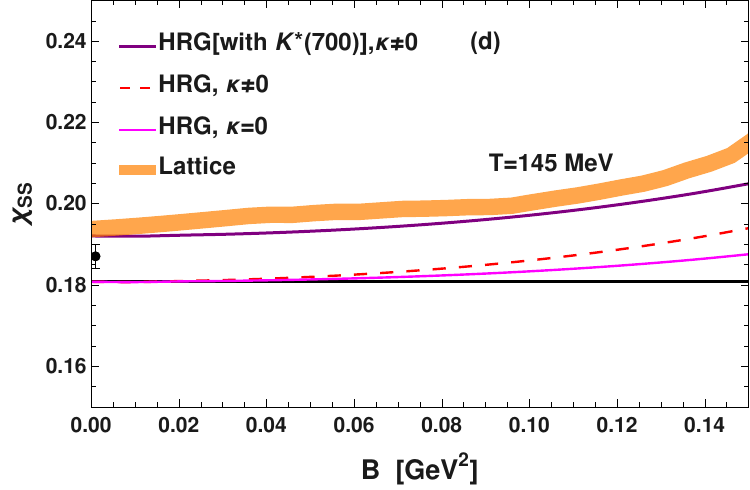} \\
\includegraphics[height=5 cm]{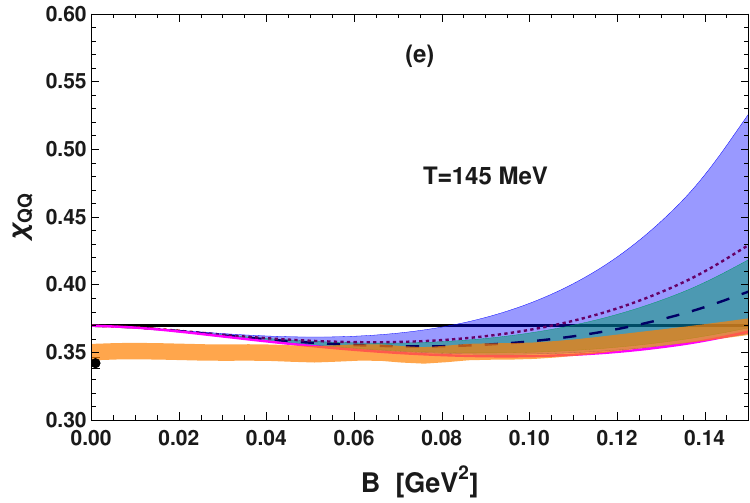}
~~~\includegraphics[height=5 cm]{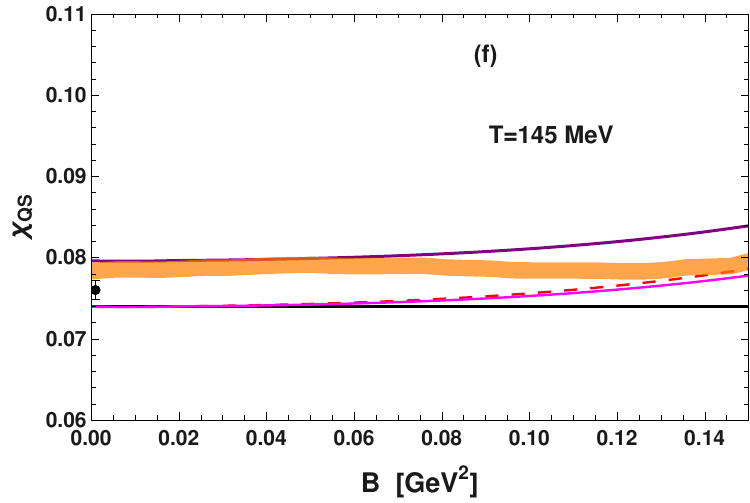} 
\end{center}
\caption{Conserved charge susceptibilities, plotted as functions of the magnetic field $B$. 
The orange bands represent the lattice QCD data digitized from~\cite{Ding:2025jfz}. The black points correspond to the data at $B=0$ from~\cite{Bollweg:2021vqf}. The horizontal black lines indicate the 
values of our model susceptibilities at $B=0$, plotted for a reference. The solid magenta lines show the HRG calculation
with the $g=2Q$ (or $\kappa=0$) prescription. The remaining bands and lines show various predictions of HRG with hadrons having the physical values of 
the $g$-factors, but for different values of $g$ for the $\Delta$ isobars, as indicated in the legends and specified in Table~\ref{tab:hadrons} (see the text for details).
This affects the plots in panels (a), (c), and (e), but not for (b), (d), or (f), where $\Delta$ does not contribute and where the HRG results with 
physical $g$-factors are drawn with dashed red lines. The solid purple lines in panel (d) and (f) represent the results with the addition of $K^\ast(700)$ states in the hadron list.  
\label{fig:suscsmallB}}
\end{figure*}

This Section presents the main results of our paper, namely, the dependence of the conserved charge susceptibilities on the magnetic field with
physical $g$-factors used in HRG. 
We restrict ourselves to sufficiently small magnitudes of the magnetic field, up to $B\sim 0.15$~GeV$^2$, which satisfies the 
approximate condition (\ref{eq:upperB}), allowing for the treatment of hadrons as structureless particles. The magnetic moments of hadrons are 
 incorporated according to the policy stated at the end of Sec.~\ref{s:hrg}.

We start with panel (a) of Fig.~\ref{fig:suscsmallB}, which presents the baryon susceptibility $\chi_{BB}(B)$. 
The legends specify various curves and bands in the figure. 
The orange bands represent the susceptibilities from the lattice~\cite{Ding:2025jfz}, which were obtained by
digitizing the plots. The additional data points at $B=0$ from~\cite{Bollweg:2021vqf} are indicated in black.
For a better reference, we draw horizontal black lines at the values of the model susceptibilities at $B=0$. 
The solid magenta lines represent the HRG calculation
with the $g=2Q$ (or $\kappa=0$) prescription, as used in~\cite{Marczenko:2024kko,Vovchenko:2024wbg}. 

The other bands and lines in Fig.~\ref{fig:suscsmallB}(a) show various predictions of HRG with the physical values of 
the $g$-factors. 
The magnetic moments ($g$ factors) for the nucleons are precisely measured~\cite{ParticleDataGroup:2024cfk}, therefore contain negligible uncertainty. The next dominant contributions comes from the $\Delta(1232)$ for which we lack a detailed determination~\cite{ParticleDataGroup:2024cfk}.
For this reason we probe different values of $g$ for the $\Delta(1232)$ isobars, as listed in the appropriate entries in Table~\ref{tab:hadrons}. The green band  in Fig.~\ref{fig:suscsmallB}(a) shows the result with the $\chi$PT values~\cite{Li:2016ezv}
with errors for the $g$ of all the $\Delta(1232)$ charge states, as given in Table~\ref{tab:hadrons}.
The blue band corresponds to the very wide range of the PDG~\cite{ParticleDataGroup:2024cfk} data for $g_{\Delta^{++}}$ (cf. Table~\ref{tab:hadrons}). As for the other 
charge states of $\Delta$, the $g$ factors are experimentally poorly known (or unknown), hence for these states we take the $\chi$PT values while using the PDG range for $\Delta^{++}$. The same recipe is used for the case of the most recent experimental value of Lopez et al.~\cite{LopezCastro:2000cv}, indicated with the dotted line. 
Finally, the dashed line, passing through the middle of the green band, uses the lattice QCD prediction~\cite{Lee:2005ds} for the $\Delta^{++}$ $g$-factor. 

We note that this kaleidoscope of HRG calculations is in reasonable agreement with the lattice data, in particular for $T=145$~MeV. 
We note that the $\chi$PT band touches the data band. Importantly, the discussed 
HRG estimates are significantly above the magenta solid line indicating the $g=2Q$ prescription~\cite{Marczenko:2024kko,Vovchenko:2024wbg}. 

For the other susceptibilities in Fig.~\ref{fig:suscsmallB}, we only provide the results for $T=145$~MeV case, which is in a better agreement with the data than for $T=155$~MeV.

In panel (c) we display $\chi_{BQ}(B)$, with a qualitative behavior and the agreement with
the data similar to panel (a), but with even wider uncertainty bands from 
the errors in the $g$ values of the $\Delta(1232)$. This is mainly due to the charge $Q=2$ 
enhancement from $\Delta^{++}$ (cf. Eq.~(\ref{eq:su1}).

In panel (e) we show $\chi_{QQ}(B)$. The interesting behavior of the model observed here is that 
when $B$ is being increased, $\chi_{QQ}$ first encounters a mild decrease, followed by an increase at larger $B$, with a dip around $B=0.10$~GeV$^2$. The decrease is due to the $s=0$ states (pions, kaons etc.) for which $\chi_{QQ}$ decreases with $B$, as implied by Eq.~(\ref{eq:chis}). 
The lattice QCD data, however, do not exhibit such a clear non-monotonicity. The discrepancy at $B=0$ follows from Fig.~\ref{fig:susczeroB}~(e).

Panels (b), (d) and (f) show the corresponding results for $-\chi_{BS}$, $\chi_{SS}$ and $\chi_{QS}$, respectively. 
Since in these cases the $\Delta$ states do not contribute, we plot only one HRG curve (red dashed line) 
with  the central values of the $g$-factors from Table~\ref{tab:hadrons}. 
In general, these quantities show a weaker dependence on the magnetic field $B$ than the previously discussed cases. 
The data for $-\chi_{BS}$ in panel (b) exhibits a non-monotonic behavior in $B$, which is not observed in the HRG model calculations.
For $\chi_{QS}$, the data seems flat compared to the model.
We note a few percent mismatch at $B=0$, which follows from Fig.~\ref{fig:susczeroB}. This can be improved for $\chi_{SS}$ and $\chi_{QS}$ as we discuss below.  

For $\chi_{SS}$ and $\chi_{QS}$, the model predictions depend on the inclusion of the $K^\ast(700)$ state, as first considered in~\cite{Friman:2015zua}, where it was shown that the attraction in the $S$-wave $\pi-K$ isospin $I=1/2$ channel is partially canceled with the repulsion in the $S$-wave $I=3/2$ channel. As a result, a net 30\% contribution of the $K^\ast(700)$ state,
treated as a wide resonance, 
is left~\cite{Friman:2015zua}. 
We apply this prescription here. The inclusion of $K^\ast(700)$ improves significantly the results for $\chi_{SS}$ and $\chi_{QS}$, 
in particular near $B=0$, as seen from Figs.~\ref{fig:suscsmallB}~(d) and (f).

To summarize this section, we note that the effect of the anomalous moments on $\chi_{BB}$, $\chi_{BQ}$, $\chi_{BS}$, and $\chi_{SS}$, when compared to the results at $B=0$ indicated with the
solid black line in Fig.~\ref{fig:suscsmallB}, is substantial (about 100\% compared to the $\kappa=0$ lines), and improves the comparison with the lattice QCD data.

\section{Effect of anomalous magnetic moments on particle multiplicities\label{s:mult}}

In the previous sections, devoted to susceptibilities (corresponding to second derivatives of the pressure with respect to chemical potentials),
we focused on comparisons of HRG with lattice QCD for a system in a thermal equilibrium and, most importantly, 
in a stationary and uniform magnetic field. 
On the other hand, the description of the effects of magnetic field on the particle yields (corresponding to first derivatives of the pressure with respect to chemical potentials) measured in ultrarelativistic nuclear collisions is more difficult for several reasons. First, during a collision the magnetic field is not static, but has a time dependence resembling a short flash
originating from the moving initial charged projectiles and the final receding spectators (for estimates of the magnitude see, e.g.,~\cite{Skokov:2009qp}), 
accompanied with a long lived but an order of magnitude weaker magnetic field generated in the process of the collision via the 
eddy currents induced in the fireball~\cite{Huang:2022qdn,Huang:2024aob}. The magnetic field is not spatially uniform~\cite{Skokov:2009qp}, 
but exhibits a profile a few fm wide. Moreover, the fireball is expanding, which further complicates the dynamics, with the Hall effect appearing in addition to the Faraday induction~\cite{Gursoy:2014aka}, leading to possible flow signatures of the magnetic field.

The strength of the magnetic field is, naturally, larger in peripheral collisions than in central collisions, where it can only result from 
fluctuations (cf.~Fig.~6 in~\cite{Skokov:2009qp}). This opens a possibility of explaining the centrality behavior of the proton to pion ratio 
measured at the LHC by the ALICE Collaboration~\cite{ALICE:2013mez,ALICE:2019hno}, where the $p/\pi^+$ ratio raises by a factor of $\sim 1.3$ 
from central to peripheral collisions. In the scenario considered in~\cite{Marczenko:2024kko,Vovchenko:2024wbg}, the effect may be explained 
if a large magnetic field ($B>0.1$~GeV) occurs in peripheral collisions.

\begin{figure}[b]
\begin{center}
\includegraphics[width=\linewidth]{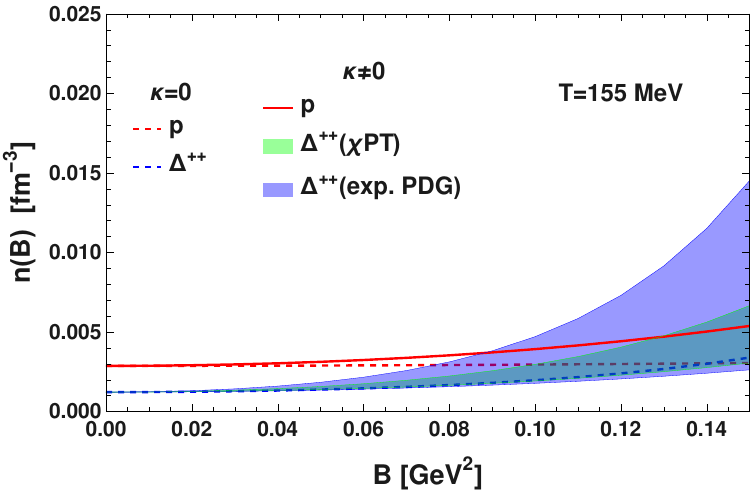}
\end{center}
\caption{Proton and $\Delta^{++}(1232)$ primordial multiplicities in  HRG, plotted as functions of the magnetic field $B$ at $T=155~{\rm MeV}$. The case without the anomalous magnetic moments ($\kappa=0$) is indicated with dashed lines while the solid lines represent the $\kappa \neq 0$ scenario. The bands for $\Delta^{++}$ correspond to the uncertainty of its anomalous magnetic moments, analogically to Fig.~\ref{fig:suscsmallB}.
\label{fig:yield}}
\end{figure} 

Here we discuss the possible role of the physical magnetic moments when amended to the argumentation of~\cite{Marczenko:2024kko,Vovchenko:2024wbg}.
For charged particles of a given species, the primordial multiplicities in thermal equilibrium at chemical freeze-out are given by the formula 
\begin{eqnarray}
\hspace{-7mm} n(B)= \frac{B|Q|}{2\pi^2} \sum_{l=0}^\infty \sum_{s_z=-s}^s \int_0^\infty dp_z \, f(B), \label{eq:n}
\end{eqnarray}
following from Eq.~(\ref{eq:pressure}). We have indicated explicitly the dependence of the distribution functions on $B$, which enters 
via the formulas for the energy spectra given in Sec.~\ref{s:hrgmag}.
A derivation analogous to the one leading to Eq.~(\ref{eq:chise}) leads, under the assumption of large $M$,  the following expansions at low $B$:
\begin{eqnarray}
&& n(B)=\frac{(T M)^{3/2}  (2 s+1) e^{-\frac{M}{T}}}{2 \sqrt{2} \pi ^{3/2}} \times \label{eq:mule}\\
&& \hspace{12mm}\left [1+ \frac{B^2 \left(g^2 s (s+1)-Q^2\right)}{24 M^2 T^2} \right ] +{\cal O} \left ( \frac{B^4}{M^2 T^2} \right ). \nonumber
\end{eqnarray}
Thus, in the large $M$ approximation,  
the leading relative dependence on $B$ is of the same form as for the susceptibilities of Eq.~(\ref{eq:chise}). We note that the relative $B^2$ term 
strongly depends on the $g$-factor and spin, while it is suppressed with $M$ and $T$. 

An exact calculation with Eq.~(\ref{eq:n}) at $T=155~{\rm MeV}$ gives the result displayed in Fig.~\ref{fig:yield}, here for the case of the proton and $\Delta^{++}(1232)$.
We note an increase of the proton primordial yield with $B$ (solid red line), with about 50\% increment at $B=0.1~{\rm GeV}^2$ compared to the value at $B=0$. This is much larger than for the case without the anomalous magnetic moment (dashed red line). 
However, the primordial protons take only $\sim 20-25\%$ of the total yield (which includes the resonance decays)~\cite{Torrieri:2004zz}. Decays of $\Delta(1232)$ supply 
another $\sim 20-30\%$, which involves a strong 
$B$ dependence (cf.~Fig.~\ref{fig:yield}), albeit with a large uncertainty. Other important feed-down contributions to 
the proton yield, including the one from $\Lambda(1115)$, have a very weak $B$ dependence. Together with the fact that the $\pi^+$ yield is very weakly dependent on $B$~\cite{Marczenko:2024kko,Vovchenko:2024wbg}, 
the anomalous magnetic moments of baryons reduce the estimate for the magnetic field needed to explain the ALICE $p/\pi^+$ data by about $10-30\%$ 
from the value $\sim 0.12~{\rm GeV}^2$ obtained within HRG in~\cite{Marczenko:2024kko,Vovchenko:2024wbg}, 
down to values $0.08-0.11~{\rm GeV}^2$. The uncertainty comes from the 
error in the $\Delta(1232)$ magnetic moments. 

Thus, even with the inclusion of the anomalous magnetic moment for the baryons, the value of the magnetic 
field needed to explain the ALICE data remains large in the light of the results of~\cite{Huang:2022qdn,Huang:2024aob}, 
brought up at the beginning of this section. Note, however, the crude nature of the approximations used, shared with~\cite{Marczenko:2024kko,Vovchenko:2024wbg}. 
In particular, the resonance feeding must be affected by the magnetic field via modification of the branching ratios, as discussed in~\cite{Marczenko:2024kko} in the context of detailed balance. Moreover, the branching ratios must depend on spin, since the 
energies of particles do depend on their spin. Therefore, a more appropriate dynamical modeling of the issue of particle yields in the presence 
of a magnetic field is desired.

\section{Summary}

We have evaluated the diagonal and non-diagonal susceptibilities related to the conserved charges in the Hadron Resonance Gas 
model in the presence of a uniform magnetic 
field in the range $0 \le B \le 0.15~{\rm GeV}^2$. A new feature is a systematic inclusion of the physical values for the magnetic moments of hadrons. 
Accounting for the anomalous parts of the magnetic moments is particularly important for the octet baryons, where the Lande $g$-factors differ very significantly from the value $2Q$ following from the minimum substitution prescription. Since the $B^2$ effects in the susceptibilities 
are roughly proportional to $g^2$, this amounts to a large enhancement. Note that the neutral baryons also contribute, due to their anomalous magnetic moments.

In our comparison to the lattice QCD data, we have included the large uncertainty of the $g$-factors of the $\Delta(1232)$ isobar (from experimental, lattice QCD, and theoretical estimates), whose contributions are particularly important for the non-strange susceptibilities
due to the large spin and isospin degeneracy. 
We have found that at $T=145~{\rm MeV}$ the agreement of the model susceptibilities as functions of $B$ with the lattice QCD data is very reasonable. Importantly, it 
improves with the inclusion of the physical $g$-factors. 
As a result, HRG with a proper list of hadron states and the proper $g$-factors is capable of reproducing the $B$-dependence of the (diagonal and non-diagonal) susceptibilities for temperatures  $\sim 150$~MeV and $0 \le B \le 0.15~{\rm GeV}^2$. For yet stronger magnetic fields, 
modification effects in the structure of the ground-state hadrons and their resonances (describing the interactions) must be incorporated. 

We have also reviewed the formulas for the energy spectra of point-like particles in magnetic moment up to spin $\tfrac{3}{2}$, used in the phenomenological studies of this paper. 

Finally, we have discussed the issue of obtaining the proton yields in ultrarelativistic nuclear collisions, taking into account the anomalous magnetic moments.

\vspace{0.5 cm}\section*{Acknowledgments}
The authors thank Michal Marczenko for valuable discussions during the revision of the manuscript. The support from the Polish National Science Center grant 2023/51/B/ST2/01625 is acknowledged.

\section*{Data Availability}
The data that support the findings of this article are openly available~\cite{Bollweg:2021vqf,Ding:2025jfz}; embargo periods may apply.

\bibliography{ref}

\end{document}